\documentclass{optica-article}

\journal{opticajournal} 

\articletype{Research Article}

\usepackage{lineno}

\usepackage{siunitx}
\newcommand{\Hz}[1]{\SI{#1}{\hertz}}

\newcommand{\nm}[1]{\SI{#1}{\nano\meter}}
\newcommand{\um}[1]{\SI{#1}{\micro\meter}}
\newcommand{\uL}[1]{\SI{#1}{\micro\liter}}
\newcommand{\mL}[1]{\SI{#1}{\milli\liter}}
\newcommand{\ms}[1]{\SI{#1}{\milli\second}}
\newcommand{\us}[1]{\SI{#1}{\micro\second}}

\newcommand{\mW}[1]{\SI{#1}{\milli\watt}}
\newcommand{\kgm}[1]{\SI{#1}{\kilogram\per\meter^3}}
 
\newcommand{\degC}[1]{\SI{#1}{\celsius}}
\newcommand{\GHz}[1]{\SI{#1}{\giga \hertz}}
\newcommand{\MHz}[1]{\SI{#1}{\mega \hertz}}
\newcommand{\kHz}[1]{\SI{#1}{\kilo \hertz}}

\usepackage{xcolor}

\begin{document}

\title{Interrogating the ballistic regime in liquids with rotational optical tweezers}

\author{Mark L. Watson,\authormark{1,2} Alexander B. Stilgoe,\authormark{1,2} Itia A. Favre-Bulle,\authormark{1,3}, and Halina Rubinsztein-Dunlop\authormark{1,2,4,*}}

\address{\authormark{1}School of Mathematics and Physics, The University of Queensland, St Lucia, Brisbane, 4072,  Queensland, Australia\\
\authormark{2}ARC CoE in Quantum Biotechnology, The University of Queensland, St Lucia, Brisbane, 4072,  Queensland, Australia\\
\authormark{3}Queensland Brain Institute, The University of Queensland, St Lucia, Brisbane, 4072, Queensland, Australia \\
\authormark{4}ARC CoE for Engineered Quantum Systems, The University of Queensland, St Lucia, Brisbane, 4072, Queensland, Australia}

\email{\authormark{*}halina@physics.uq.edu.au} 

\begin{abstract*} 
Accessing the ballistic regime of single particles in liquids remains an experimental challenge that shrouds our understanding of the particle--liquid interactions on exceedingly short time scales. We demonstrate the ballistic measurements of rotational probes to observe these interactions in the rotational regime within microscopic systems. This study uses sensitive high-bandwidth measurements of polarisation from light scattered by orientation-locked birefringent probes trapped within rotational optical tweezers. The particle--liquid interactions in the ballistic regime are decoupled from the optical potential allowing direct studies of single-particle rotational dynamics. This enabled us to determine the dissipation of rotational inertia and observe and validate rotational hydrodynamic effects in a previously inaccessible parameter space. Furthermore, the fast angular velocity thermalisation time enables calibration-free viscometry using less than \ms{50} of data. This methodology will provide a unique way of studying rotational hydrodynamic effects and enable ultra-fast microrheometry in systems out-of-equilibrium. 
\end{abstract*}

\section{Introduction}
Complex systems with rapidly changing spatiotemporal dynamics underpin biological processes \cite{Parry2014, Gnesotto2018, Junger2018, Rohrbach2020, Scott2023}, fluid dynamics \cite{Chacko2018, SaintMichel2018, Ovarlez2020}, fundamental thermodynamic laws \cite{Rings2010, Geib2020, Mayer2023}, and, more generally, systems out-of-equilibrium \cite{Argun2016, Doostmohammadi2016, Oliveira2019, Kralj2024, Viot2024}, yet there is a lack of sufficient measurement techniques to elucidate the physical and statistical properties governing them. Ballistic tweezers provides a unique tool to exploit particle--liquid interactions in microscopic systems to gain new insight into such properties \cite{Huang2011, Kheifets2014, Madsen2021}. Sensing both the translational and rotational inertia is crucial to fully capture these interactions and interrogate systems out-of-equilibrium on exceedingly short timescales. The rotational ballistic motion in microscopic systems has remained a largely theoretical research topic and requires experimental techniques to corroborate results. In this work, we developed a technique that improves upon previous rotational optical tweezers methods and gains the capability to probe the complete transition from diffusive to ballistic rotational dynamics. 

Optical tweezers (OTs) are a versatile tool in applied and fundamental physics as they can exert and sense positions, forces and torques in nano- and micro-scale systems with great precision. The technique allows for direct manipulation of a system and to extract its physical characteristics through the use of optically trapped probes. This has led to notable success in studies of the mechanical properties governing complex biological processes. This includes DNA mechanics \cite{Smith1996, Wang1997, Sheinin2011, Peddireddy2024}, cellular adhesion forces \cite{Singer2003, Fallman2004}, single molecule sensing \cite{Norregaard2017, Bustamante2021}, and cell viscoelasticity \cite{Mas2013, RobertsonAnderson2018, Watson2022, Lu2020, Stilgoe2024, Vos2024}, to cite a few. 

The physical properties of the environment, including viscosity, can be investigated by analysing the trajectory of an optically trapped probe, providing excellent spatially and temporally resolved microrheology measurements. Several well-established optical tweezers methods exist that rely on analysing the stochastic dynamics of the probe \cite{BergSorensen2004, Brau2007, Preece2011, Mas2013, Zhang2017, Neckernuss2016, Berghoff2021}. However, OTs can also access energy scales beyond stochastic dynamics and probe the transition into Newtonian dynamics as the temporal resolution improves \cite{Huang2011, Kheifets2014, Madsen2021}. This opens a new measurement space to study particle--liquid interactions, specifically those on exceedingly short time scales down to milliseconds. 

In liquids, micron-sized particles move due to thermal motion where kinetic energy is gained from random collisions with fluid molecules. The kinetic energy quickly dissipates giving rise to the apparent instantaneous displacements observed in Brownian motion. This is defined as the diffusive regime. However, the dissipation of the particle's velocity can be observed with sufficient spatial and temporal resolution allowing access to the ballistic regime where the velocity can be resolved. Achieving this resolution is limited by the signal-to-noise ratio (SNR) of the detection system. However, improved SNR can be achieved using optically based methods. 

Optical tweezers routinely interrogates the diffusive regime, and more recently, have successfully resolved the ballistic regime \cite{Huang2011, Kheifets2014, Madsen2021} in translational geometry in liquids. Trapping experiments are performed most commonly in the diffusive regime and measure the position relaxation of the probe. This requires stiff optical traps and measurement times on the order of seconds. Improving the sensitivity to access the ballistic regime and instead measuring the velocity dissipation of the probe allows for unique studies of particle--liquid interactions on a spatiotemporal scale that was previously inaccessible. This is particularly important in the context of mobility and microrheometry, which differ between timescales \cite{Brau2007, Preece2011, Zhang2017}. High-bandwidth measurements in liquids have resolved translational ballistic motion, enabling validation of hydrodynamic effects \cite{Huang2011,Kheifets2014} and ultrafast viscometry measurements \cite{Madsen2021}. However, to fully characterise a system and understand the complete transition between continuum and stochastic mechanics, it is also crucial to elucidate the rotational dynamics. We demonstrate measurements in the rotational ballistic regime, where the dissipation of angular velocities is observed. 

Rotation can be introduced to the system by transferring optical angular momentum from light to the probe in addition to the linear momentum transfer required for trapping, giving birth to Rotational Optical Tweezers (ROTs) that is capable of exerting and precisely measuring torques \cite{Bishop2004, LaPorta2004,  Bennett2013, Zhang2017, Roy2018, Lokesh2021, Strasser2022}. Typically, ROTs use a birefringent microsphere as a probe to transfer spin angular momentum from light. There is a measurable change in the polarisation state of the trapping beam, which provides a useful and precise method to sense both torques and angular displacements. A rotating probe in ROTs does not move far from the centre of the trap. This enables studies of DNA twisting mechanics \cite{Ma2013}, adhesion \cite{Knoner2006, Vaippully2019}, and microrheometry \cite{Bishop2004, LaPorta2004, Shreim2012, Bennett2013,  Zhang2017, Stilgoe2024}. Recently, ROTs were used for intracellular viscometry \cite{Vaippully2020, Watson2022, Roy2023}, highlighting their unique applicability in confined spaces or near surfaces \cite{Leach2009, Zhang2019}.

Rotational measurements based on polarisation detection have inherent advantages over the typical deflection-based techniques employed in translational measurements. The latter responds to changes in the outgoing light and is most greatly affected by interactions at the beam focus. Unfortunately, deflected photons make up a small portion of the collected light, limiting the SNR in the detection and obscuring the ballistic regime. The signal can be improved by increasing the trapping power, using high-refractive index particles, or by interferometric methods that filter out the trapping beam \cite{Kheifets2014, Madsen2021}. However, these methods have increased complexity compared to traditional measurements which limits their use. 

The detection of angular position and torque using polarisation can achieve a greater SNR as each photon carries information on the beam polarisation state. The SNR is limited by the amount of spin angular momentum transferred to the particle (which is determined by the birefringence) and the noise in the detection apparatus. By introducing a balanced photodetector to the measurement system---that cuts out common mode noise---we show that ROTs become sensitive enough to detect the transition into rotational ballistic dynamics. Improved noise characteristics in our experiments allow information to be obtained from analysis of the angular velocity power spectrum (AVPS) \cite{Madsen2021}---which has not been possible in our previous rotation experiments. This enhanced capability enables measurements of particle--liquid interactions on extremely short-time scales.

\section{Rotational Ballistic Regime}
Consider a spherical birefringent probe that is optically trapped in a purely viscous liquid. The rotational and translational components are decoupled due to the symmetry of the system; hence, we can examine the rotational dynamics alone. Vaterite microspheres are excellent high birefringence optical probes (calcium carbonate, positive uniaxial birefringent crystals with $\Delta n \approx 0.1$) that can be reliably synthesised with good control over their morphology \cite{Vogel2009, Vikulina2021}. The interaction of a linearly polarised beam with their birefringence produces an optical torque that aligns the extraordinary axis of the probe with the polarisation state of the incident beam. At small displacements, near equilibrium, this can be approximated as a harmonic potential \cite{Bennett2013}. As vaterite possesses positive uniaxial birefringence, circularly polarised light causes the trapped probe to rotate at a constant rate limited by the drag---with Stokes drag coefficient, $\gamma_0 = 8\pi\eta a^3$, where $\eta$ is the viscosity of the fluid and $a$ is the radius of the probe.  

\begin{figure}[ht]
    \centering
    \includegraphics[width=0.75\columnwidth]{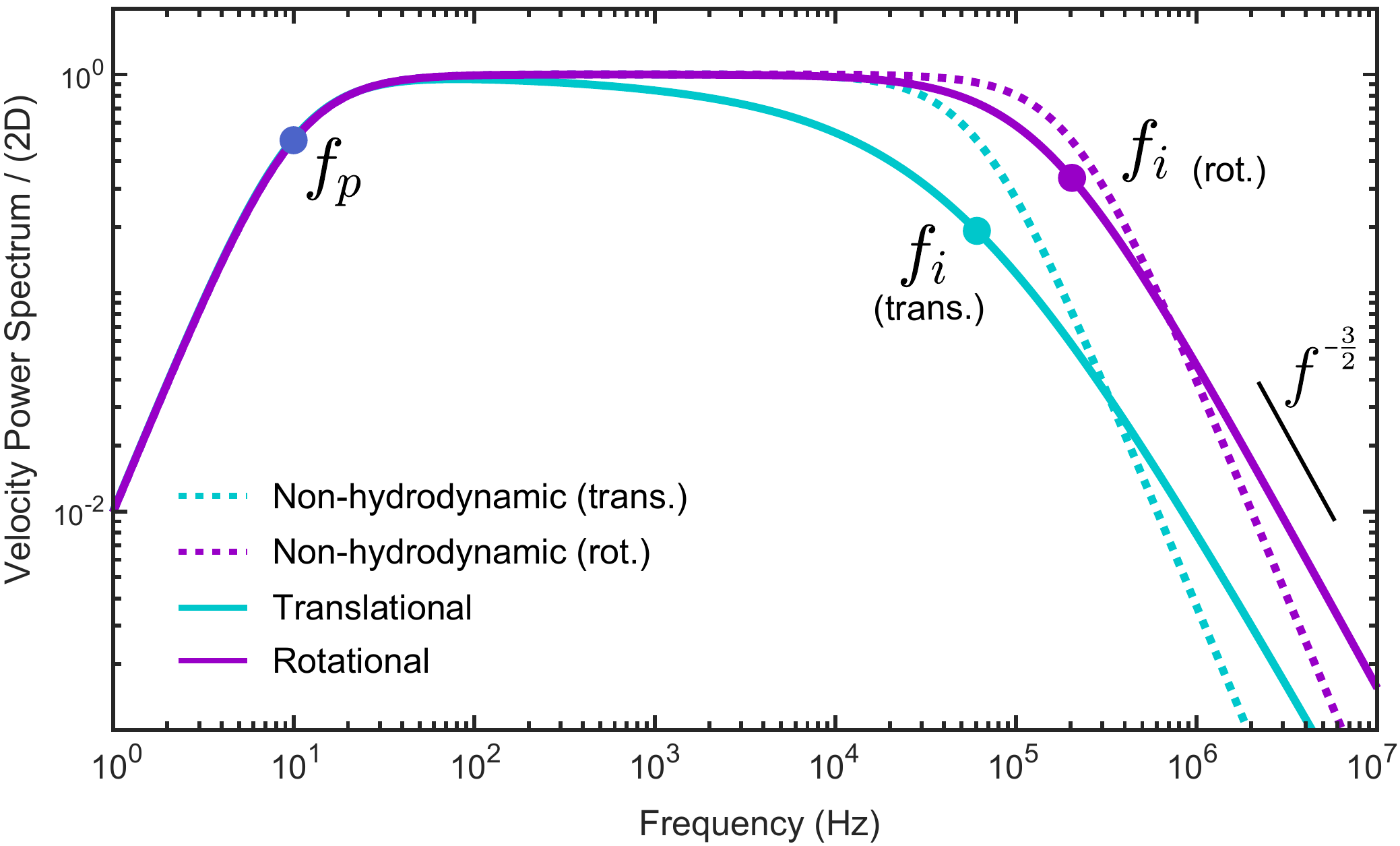}
    \caption{The translational (cyan) and rotational (purple) velocity power spectra for a \um{2} radius microsphere normalised by their respective diffusion constant and with $f_p=$ \Hz{10},  $f_{i\mathrm{,trans.}}\approx$ \kHz{61}, and $f_{i\mathrm{,trans.}}\approx$ \kHz{202}. The dashed lines indicate the models without hydrodynamic modifications (equivalent to $\gamma(f)\rightarrow\gamma_0$ in Eq. \ref{eqn:VPS}) for each case. The black line indicates a slope $\propto f^{-3/2}$. }
    \label{fig:VPScomparison}
\end{figure}

Thermal torque impulses lead to stochastic motion that is opposed by the position-dependent optical torque and the velocity-dependent drag torque. The equation of motion that results is given by $I\ddot{\theta}(t) = T_\text{optical} + T_\text{thermal}+T_\text{drag}$, where $I$ is the moment of inertia, and $T$ represents the torques in the system. As the reactionary optical torque aligns the probe to an equilibrium orientation, a characteristic timescale, $\tau_p$, emerges associated with the angular position relaxation. Similarly, the drag torque from friction with the fluid opposes the angular velocities and a second characteristic timescale, $\tau_i$, emerges relating to the dissipation of rotational inertia. At low Reynold's numbers $\tau_i\ll\tau_p$ which causes the inertial term to be neglected in most OTs experiments. These two parameters introduce corners into the angular velocity power spectrum at frequencies of $f_p=2\pi/\tau_p$ and $f_i=2\pi/\tau_i$ respectively, shown in Fig. \ref{fig:VPScomparison}. 

At short timescales nearing $\tau_i$, the probe's motion is complicated by the hydrodynamic interaction with the fluid and cannot be treated as linearly proportional to velocity. Instead, the drag depends on the past motion of the probe and results in a frequency-dependent drag coefficient, $\gamma(f)$. The rotational drag for a rotating sphere is given by,
\begin{equation}
    \gamma(f) =\gamma_0\left( 1+\frac{2}{3}\frac{(1-i)(f/f_v)^{3/2} - i(f/f_v)}{1+2(f/f_v)^{1/2} + 2(f/f_v)}\right),
    \label{eqn:drag_rot}
\end{equation}
where we define a new characteristic frequency, $f_v$, associated with the dissipation rate of the fluid's inertia to parametrise the equation. The corresponding timescale, $\tau_v=2\pi/f_v$, is of the same order of magnitude as $\tau_i$ and can be defined in terms of $f_i$. When $f\ll f_v$, $\gamma(f)$ approaches the Stokes drag constant for a rotating sphere at low Reynold's numbers, $\gamma_0$. The inertia of the fluid envelope surrounding the probe modifies the mass term to an effective mass that includes the displaced fluid such that $m = m_\mathrm{particle} + m_\mathrm{fluid}/2$ \cite{Zwanzig1975, Kheifets2014}. Hence, $I=I_\mathrm{particle} + I_\mathrm{fluid}/2 = (2/5)ma^2$ becomes an effective moment of inertia. Compressibility effects are neglected because they dissipate significantly faster than the measurement rate \cite{Kheifets2014, Madsen2021}.

The transition between diffusive and ballistic behaviour is best observed in the Angular Velocity Power Spectrum, AVPS$=(2\pi f)|\Tilde{\theta}(f)|^2$, where $\Tilde{\theta}(f)$ is the unilateral Fourier transform of the angular displacement from equilibrium, $\theta(t)$. The velocity power spectra (VPS) for the translational and rotational cases are shown in Fig. \ref{fig:VPScomparison} and can both be described by,
\begin{equation}
    \text{VPS}=\frac{2D \text{ Re} \left[ \gamma(f)/\gamma_0 \right]}{\left(f_p/f - f/f_i + \text{ Im}\left[\gamma(f)/\gamma_0\right]\right)^2 + \left(\text{Re} \left[ \gamma(f)/\gamma_0 \right] \right)^2},
    \label{eqn:VPS}
\end{equation}
with appropriate substitutions of each parameter for the two cases. These parameters are defined in Table \ref{table:VPSparameters} for both translational and rotational motion. The translational $\gamma(f)$ is not listed in this table but the function is well established \cite{BergSorensen2004, Madsen2021}. The characteristic frequencies of inertia dissipation in the translational and rotational cases are related by $f_{i\mathrm{,rot.}}=(10/3)f_{i\mathrm{,trans.}}$. The similar magnitudes imply a similar level of sensitivity is required for resolving ballistic motion in both cases. 

\begin{table}[hbt]
    \centering
    \begin{tabular}{ccllll}
        Parameter   &  Symbol  && Translational  && Rotational \\  \hline 
        Stokes Drag &{$\gamma_0$} && $6\pi \eta a$   && $8\pi\eta a^3$  \rule{0pt}{2.6ex} \\
        Diffusion &{$D$} && $k_BT/\gamma_0$   && $k_BT/\gamma_0$ \\
        Position Relaxation &{$f_p$}   && $\kappa/2\pi\gamma_0$  && $\chi/2\pi\gamma_0$ \\
        Inertia Dissipation&{$f_i$}   && $\gamma_0/2\pi m$ && $\gamma_0/2\pi I$ \\ 
        Flow Relaxation&{$f_v$}   && $\gamma_0/9\pi m_\mathrm{fluid}$ && $\gamma_0/15\pi I_\mathrm{fluid}$
    \end{tabular}
    \caption{The parameters in the translational and rotational velocity power spectra (Eq. \ref{eqn:drag_rot}-\ref{eqn:VPS}), defined in terms of the viscosity ($\eta$), radius ($a$), linear ($\kappa$) and angular ($\chi$) trap stiffnesses, Boltzmann's constant ($k_B$) and temperature ($T$).}
    \label{table:VPSparameters}
\end{table} 

Fig. \ref{fig:VPScomparison} compares the linear and angular velocity power spectra and highlights the relative hydrodynamic contribution in each. The spectra have been normalised by their respective diffusion constant and the position relaxation frequency has been set to $f_p=$\Hz{10}. There is a negligible difference in the low-frequency limit (diffusive regime) as both cases have harmonic potentials and the drag is linearly proportional to the velocity. In the high-frequency limit (ballistic regime), the spectra decay proportionally to $f^{-3/2}$ due to the modified drag term. Vortices are formed in response to the motion of the probe and decay at a rate related to $f_v$. Although $f_v$ is similar in magnitude for both cases, linear displacements generate vortices in the surrounding fluid, whereas angular displacements generate a line of vortices along the rotation axis originating within the probe. The vortices in the fluid propagate further and take longer to decay leading to greater hydrodynamic effects observed in the translational case between $f_p$ and $f_i$. Ballistic tweezers have experimentally verified the translational drag model for micron-sized particles in liquids \cite{Kheifets2014, Madsen2021}, whereas the rotational analogue with its significantly smaller deviation has yet to be investigated. The rotational ballistic measurements described in this study provide a viable technique for experimentally verifying these dynamics. 

\section{Materials and Methods}
\subsection{Rotational Optical Tweezers}
\begin{figure}[ht]
    \centering
    \includegraphics[width=1\linewidth]{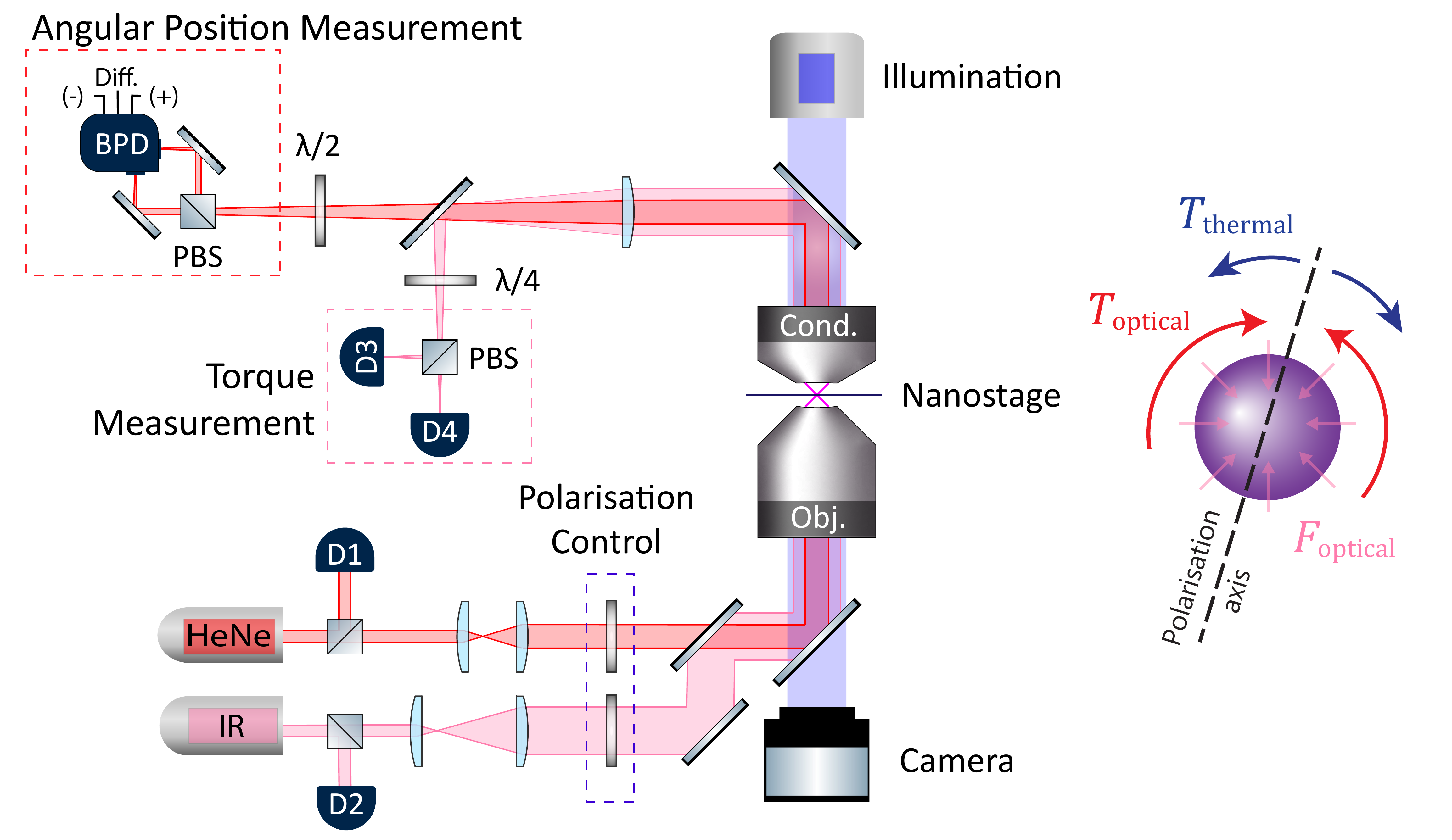}
    \caption{Simplified schematic of the rotational optical tweezers setup, adapted from \cite{Watson2022}. Photodetectors D1 and D2 (PDA36A-EC, Thorlabs) measure the initial power of each beam. The beams are expanded and a series of waveplates controls the incident polarisation before entering the high NA objective to form the optical trap in the sample. The outgoing light is collected by a condenser where D2 and D3 measure the circular polarisation states of the IR beam to determine the optical torque and the balanced photodetector (BPD) measures the linear polarisation state of the He--Ne beam to determine angular positions. The BPD outputs the corresponding voltage (+ and - output channels) and an amplified voltage signal of the difference between them (diff.). The right illustrates the force and torque a vaterite experiences in a linearly polarised trap where the beam axis is coming out of the page. }
    \label{fig:ROTsetup}
\end{figure}

The optical trapping component of the set-up has been described previously  \cite{Zhang2017,Watson2022} and a simplified schematic is depicted in Fig. \ref{fig:ROTsetup}. Briefly, trapping is achieved using a \nm{1064} NdYAG laser with powers at the trap between 20-\mW{200}. A half waveplate controls the equilibrium orientation of the vaterite when the trap is linearly polarised. A removable quarter waveplate enables switching between linearly and circularly polarised traps. The degree of circular polarisation of the forward scattered light is measured to calibrate the system and determine the optical torque \cite{Watson2022}.

A low-powered (<\mW{5}) and weakly focussed Helium--Neon laser (HL210LB, Thorlabs) measures angular displacements of the vaterite. The He--Ne beam provides an independent measurement basis with negligible effects to the optical trap and has a superior noise spectrum compared to our trapping beam such that it is capable of resolving a signal at higher frequencies. The outgoing linear polarisation states are measured using a balanced photodetector (HBPR-100M-60K-SI, Femto). The signal from the photodetector is captured using an oscilloscope (DHO4240, Rigol) at \GHz{1} to minimise the noise contribution from the device and ensure a sufficient acquisition rate to avoid aliasing in the measurement bandwidth.

Introducing the balanced photodetector enables access to higher frequency measurements previously unobtainable in ROT experiments. As opposed to typical photodetectors that output a voltage proportional to the intensity of the light, the balanced photodetector (BPD) measures the difference in intensity between two sources. This is ideal for measuring the signal from the orientation-locked vaterite which causes small changes to the polarisation as it fluctuates under rotational thermal motion. In our set-up, a polarising beam splitter separates the light into orthogonal linear polarisation states that are each incident on the BPD. Measuring the difference between intensities suppresses the common-mode noise up to 40dB and effectively gains an order of magnitude in temporal resolution compared to previous measurements. The maximum sensitivity occurs when the two states have equal powers and this is achieved by orientating the vaterite such that the polarisation states are equal. Shot noise will vary slightly between the two detectors which places an upper limit on the effectiveness. However, the SNR in these experiments was limited by laser noise and vaterite morphology.

\subsection{Sample Preparation}
Vaterite microspheres were synthesised according to the procedure outlined by \cite{Vogel2009}. Vaterite is a positive uniaxial birefringent crystal with a density of \kgm{2540} which has an approximately hyperbolic internal structure when synthesised as a sphere \cite{Parkin2009} with refractive indexes of $n_o=1.55$ and $n_e=1.64$. Briefly, \SI{1}{\mol} of aqueous solutions of CaCl$_2$ (\mL{5}), MgSO$_4$ (\mL{2}), and K$_2$CO$_3$ (\mL{1.5}) were mixed for three minutes followed by the addition of \uL{50} AGFA Agepon to stabilise the reaction and prevent further crystal growth. The vaterites were washed three times with ethanol. There are a variety of methods for controllable synthesis of vaterites with size ranges varying from \um{0.5} to \um{25} in diameter \cite{Parakhonskiy2012, Christy2017, Trushina2016, KonopackaLyskawa2019, Vikulina2021}; however, much of this research does not consider the birefringence of vaterites, rather focus on their capability for drug loading and release. Our synthesis method enabled us to produce vaterite microspheres with a highly reproducible size range and birefringence, which are the key parameters for ROTs experiments. 

\uL{50} of vaterite suspension was added to \mL{1} of the medium of interest, which for the measurements included in this work, was water and water--glycerol solutions. The solution was sonicated to separate any vaterite microspheres that clumped together. Care must be taken to ensure the concentration of the system is not too dense to avoid trapping multiple particles.  

\section{Results and Discussion}
\subsection{Accessing the Ballistic Regime}

\begin{figure*}[hbt]
    \centering
    \includegraphics[width=\columnwidth]{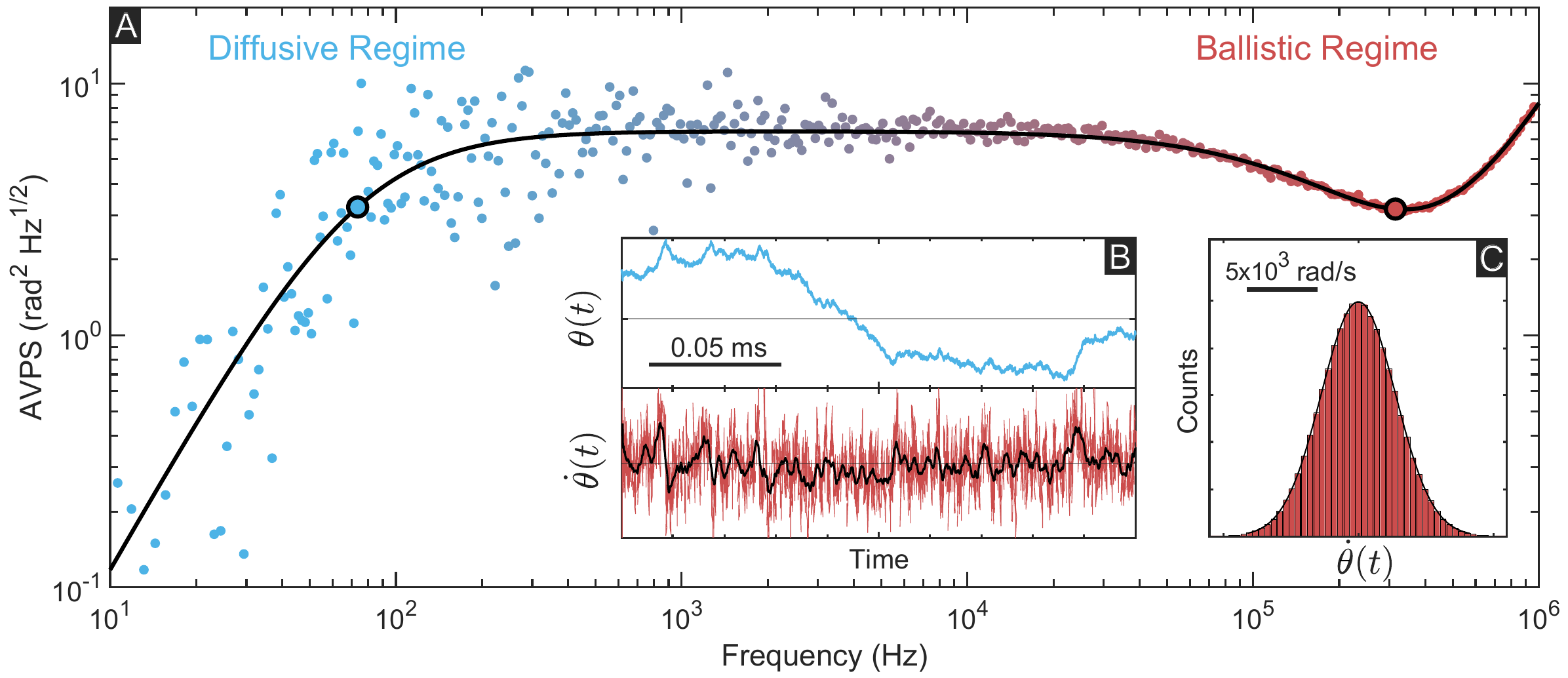}
    \caption{(A) Wideband AVPS measurement for a \um{2.3} sized probe measured at \MHz{10} for \SI{1}{\second} showing the transition from the diffusive to the ballistic regime. The data was fitted to Eq. \ref{eqn:VPS} (black line) with an empirically determined noise floor and used to determine $f_p$ and $f_i$ (blue and red circles on the curve). (B) A \ms{2} snapshot of the measured angular position, $\theta(t)$, and angular velocity, $\dot{\theta}(t)$, time traces for \um{1.4} radius probe measured over \ms{50} with the data low pass filtered at \MHz{10}. Further windowing of $\dot{\theta}(t)$ with \us{5} is shown in the black line. (C) A histogram of $\dot{\theta}$ captured over the full \ms{50} and fitted with a normal distribution (black line).}
    \label{fig:BallisticRegime}
\end{figure*}

The AVPS in Fig. \ref{fig:BallisticRegime}A demonstrates the full transition between the rotational diffusive and ballistic regime. It is experimentally observed as evidenced by the two corners in the curve becoming resolvable at the characteristic frequencies of $f_p$ and $f_i$. The data is fitted to the AVPS model in Eq. \ref{eqn:VPS} with fitting parameters for $D$, $f_P$, and $f_i$, as well as an additional noise term characterised empirically from the signal measured from an empty trap. The magnitude of the noise floor varies due to acquisition settings and the birefringence of the vaterite. For this wideband measurement that captures the full transition, the signal becomes dominated by noise at frequencies above $f_i$. Our typical high-frequency measurements can resolve the power spectra beyond \MHz{1} as they only determine $f_i$ and not $f_p$, as discussed in the following section. 

Importantly, the spectrum plateaus between $f_p$ and $f_i$ as predicted by the hydrodynamically-correct model shown in Fig. \ref{fig:VPScomparison}. This indicates that the signal from our rotational measurement is decoupled from the translational dynamics as translational hydrodynamic contributions would cause a deviation at lower frequencies. 

The high-frequency measurements aimed at determining $f_i$ and were measured at \MHz{100} for \ms{50}. The short measurement time prevents the diffusive regime and $f_p$ from being resolved but provides sufficient time to capture the rotational ballistic dynamics and measure angular velocities. Fig. \ref{fig:BallisticRegime}B shows a \ms{2} time trace of the observed angular position and velocity of the probe. The increased sensitivity of our detection systems means $\theta(t)$ no longer appears as a stochastic signal and instead becomes differentiable enabling the angular velocity to be determined, $\dot{\theta}(t) \approx \Delta\theta/\Delta t$. The measured signal is low-pass filtered at \MHz{10} to remove the noise-dominated spectrum which reveals that the slope $\theta(t)$ is correlated with $\dot{\theta}(t)$. The angular velocity thermalises to reveal a Maxwell-Boltzmann distribution demonstrated in Fig. \ref{fig:BallisticRegime}C. The measured $v_{rms} = (2.744\pm0.004)\times10^3$~\unit{\radian^2 \per \second^2} reasonably agrees with the expected modified-equipartition theorem with $v_{rms}\approx 2.88\times10^3$~\unit{\radian^2 \per \second^2}. Observing the correlation in the angular velocity trace and its distribution is an important step in verifying a successful ballistic measurement \cite{Huang2011}. 

Vaterites are useful and convenient optical probes frequently used in ROTs experiments as they can be easily synthesised, are close to spherical, and have strong birefringence ($n_e = 1.65$, $n_o=1.55$, $\Delta n\approx 0.1$). Birefringence is important for generating optical torque and achieving high SNR because it causes the spin angular momentum transfer from the light and thus affects the sensitivity of the polarisation-dependent measurements of angular position. We observed a variation in the birefringence between otherwise equal vaterite probes which changes the resolving power of the angular displacements. Vaterite has low absorbance at \nm{633} and \nm{1064} which is crucial to minimise localised heating around the probe which would otherwise change the local fluid properties.

Probes below \um{0.5} radius can become less spherical and slightly more difficult to trap, which sets a reasonable upper limit for the required sensitivity of our technique ($f_i=1.74$ \MHz{} in water at \degC{20}). Vaterites in ROTs experiments are more commonly sized 1-\um{3} in radius with corresponding inertia dissipation rates of 0.44-\MHz{0.04}. The spherical geometry simplifies the equations of motion and helps to decouple the translational and rotational degrees of motion. It should be noted that spherical geometry is less important for measurements aimed at sensing optical torque as the rotational dynamics, including the hydrodynamic effects, can be neglected. Variation in shape could cause coupling of translational and rotational dynamics. However, the aspect ratios of particles used in our experiments were close enough to unity so that coupling was small. The flat plateau in the AVPS curve in Fig. \ref{fig:BallisticRegime}A implies negligible translational-rotational coupling. This is strong evidence that our detection technique is sensitive only to rotation. 
 
There are few alternatives to vaterite as a probe in our system, however, the possibility of engineering probes with controllable birefringence and size remains a tantalising opportunity to improve the sensitivity and precision of rotational ballistic measurements. Vaterite microspheres can also be coated and functionalised \cite{Vogel2009}. This improves their stability in more environments and can be very beneficial when applied to a large variety of biological systems. 

\subsection{Ultrafast Measurements}
\begin{figure}[ht]
    \centering
    \includegraphics[width=1\linewidth]{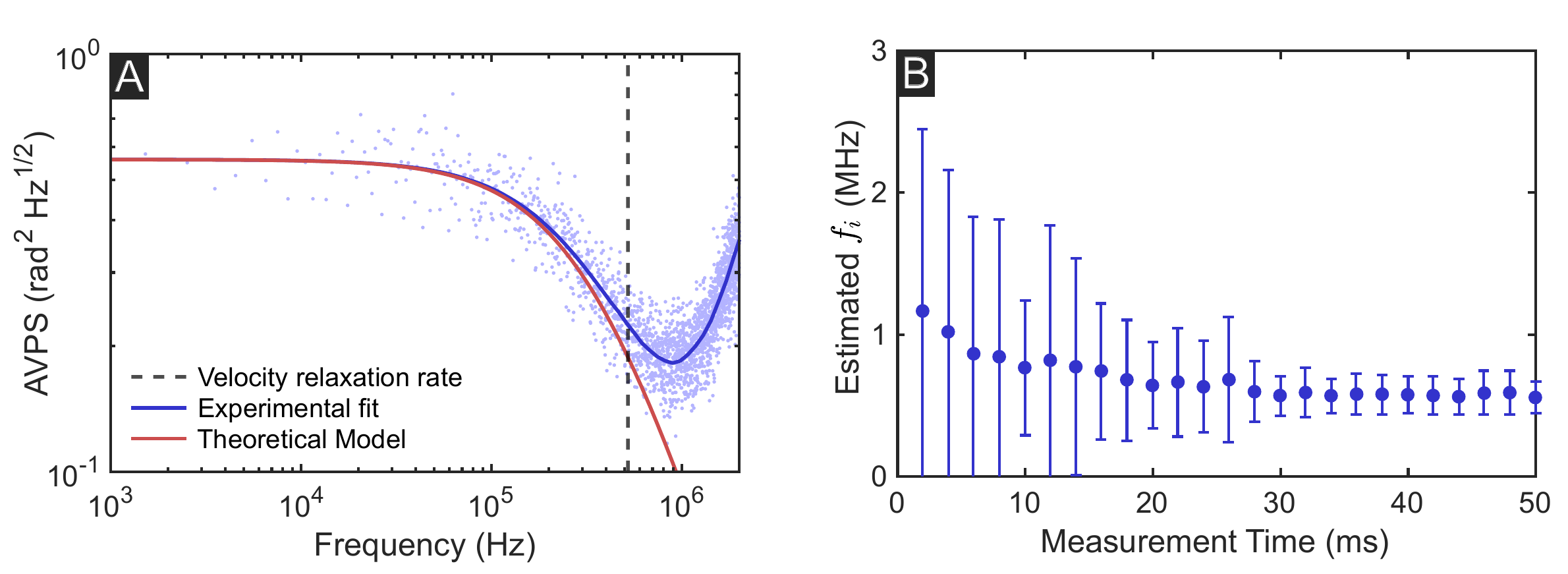}
    \caption{(A) A typical experimental AVPS (blue dots) in water overlaid with the theoretical model (red line). The data has been windowed into \kHz{1} bins and the dashed line indicates the determined angular velocity relaxation rate, $f_i$. (B) The estimated value of $f_i$ varying measurement times that are short segments from a \ms{100} sample.}
    \label{fig:AVPS_ballistic}
\end{figure}

Ballistic measurements were performed with \ms{50} measurements times captured at \MHz{100}. The data was windowed into \kHz{1} bins and truncated at high frequencies once the signal-to-noise (SNR) $>1$ and information can no longer be retrieved. Non-linear Least-squares fitting of the hydrodynamically-correct AVPS model in Eq. \ref{eqn:VPS} was carried out. As the total measurement time was too short to resolve the angular position relaxation, only $D$ and $f_i$ were used as free parameters with $f_p$ initially determined from longer measurements. The noise profile of the system was characterised by calibration measurements without a particle in the optical trap and scaled to the optical power on the detector. Fig. \ref{fig:AVPS_ballistic}A shows a typical experimental AVPS fitted with the blue line and the calculated value of $f_i$ at the frequency indicated by the dashed line. The deviation from the expected model in red was caused by the signal approaching the noise floor ($\propto f^2$) which became dominant beyond \MHz{1}. The shape of the noise floor is easily characterised however its magnitude and hence its prevalence depends on the SNR and the optical power on the detector. The SNR changes slightly between probes due to variations in morphology and birefringence which alters the amount of spin angular momentum that can be transferred from the beam. 

As the hydrodynamic effect has a negligible contribution to the AVPS below \kHz{10}, $2D$ can be estimated by averaging those values. There was a maximum error of 6\% between the predicted value of $D$ and the fitted value providing an accurate estimation for the diffusion constant. Similarly, $f_i$ was predicted from $\text{AVPS}|_{f=f_i}\approx D/1.49$ and was within $30\%$ of the fitted value. Measurements where the value of $f_i$ is determined with SNR$<1$ cannot be resolved adequately and were removed from the analysis. These limitations can be attributed to specific properties of the vaterite probes in those cases, mainly decreased birefringence, which disrupted the effectiveness of the polarisation-sensitive measurements ($<10\%$ of probes). As $f_v$ can be defined in terms of $f_i$ along with knowledge of the densities, the hydrodynamic contributions were also parametrised with $f_i$ ensuring only two fitting parameters are needed. The agreement of the fit experimentally verifies the rotational hydrodynamic model of a single microsphere in a viscous fluid in a new measurement bandwidth. The technique provides a unique opportunity to validate rotational hydrodynamic models approaching the \MHz{} region for micron-scale systems. 

The SNR depends on the sensitivity in resolving angular displacements, which is largely determined by the birefringence of the vaterite probe. An ideal probe for our polarisation-based detection is a strongly birefringent microsphere with low absorbance and reflectance. The spherical geometry simplifies the hydrodynamic model, while the latter properties are required to generate an alignment torque and measure the orientation from the polarisation changes in the forward scattered light. Since the aim of rotational ballistic measurements is to resolve the angular velocity dissipation rate, lowering $f_i$ away from the noise-dominated high-frequency region by increasing the probes radius or lowering the viscosity achieves this goal.

A key advantage of ballistic measurements is the short total measurement time required to retrieve a measurement of $f_i$. Fig. \ref{fig:AVPS_ballistic}B shows the estimated value of $f_i$ for increasing measurement times from the same experiment. Although the data is windowed into \kHz{1} bins, measurement times longer than \ms{1} are necessary to obtain sufficient time-averaging to adequately resolve the AVPS for fitting. A minimum measurement tie of \ms{30} was required to constrain the uncertainty of $f_i$ within $10\%$. This minimum time varies slightly between probes based on their size and birefringence; however, we determined \ms{50} of measurement time is sufficient to resolve $f_i$ for all of our appropriate $2$-\um{4} sized probes.  

\begin{figure}[ht]
    \centering
    \includegraphics[width=0.9\linewidth]{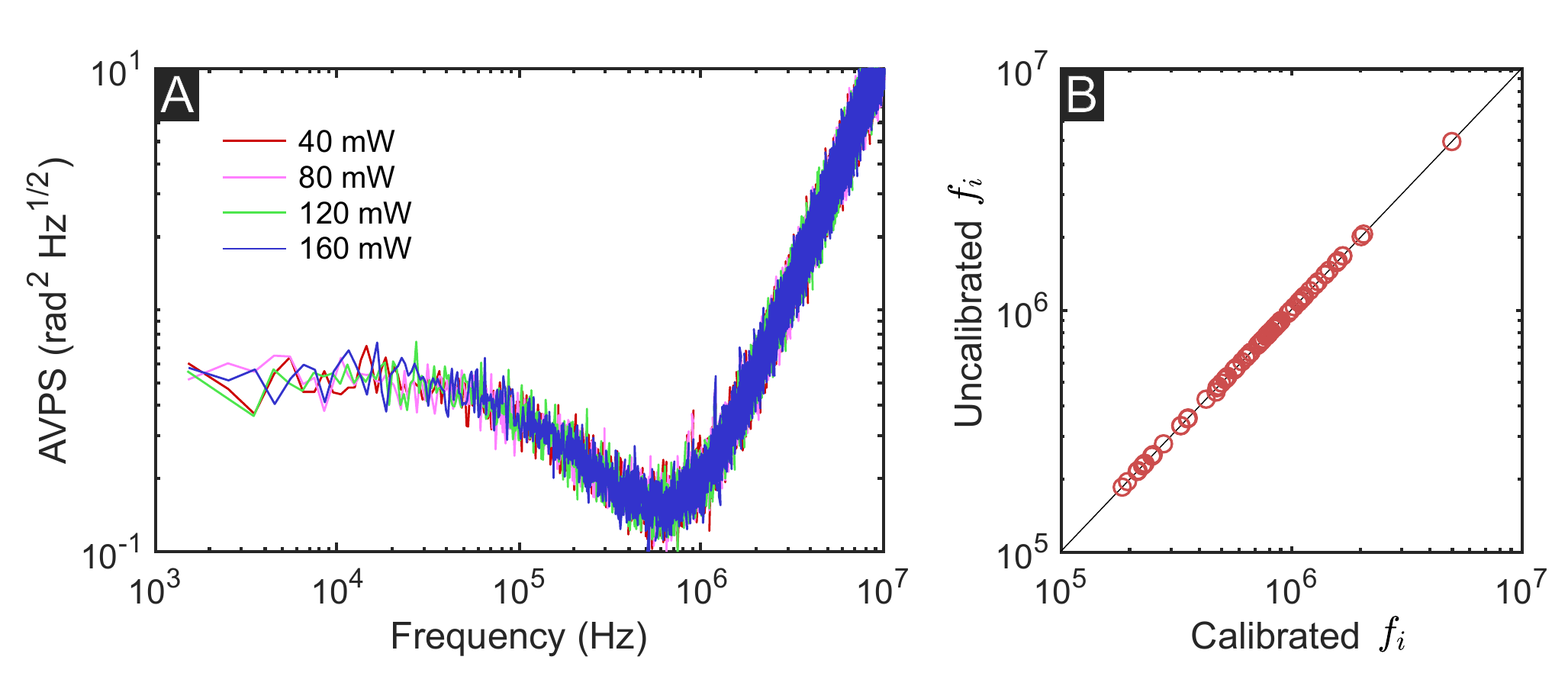}
    \caption{(A) Experimental AVPS captured for the same probe in water with different trapping powers. (B) The estimated values of $f_i$ from fitting with a calibrated value of $f_p$ compared to an uncalibrated model with $f_p=0$. There was a mean difference of $0.07\%$ between the models.}
    \label{fig:CalibrationFree}
\end{figure}

The angular position trajectory in the ballistic regime is decoupled from the optical potential which provides a major advantage for ballistic tweezers applications because the position relaxation time, $\tau_p$, is significantly larger than the total measurement time. Hence, the optical trap in the AVPS has a negligible effect once $f\gg f_p$. Fig. \ref{fig:CalibrationFree}A shows the AVPS from repeated ballistic tweezers measurements in water with trapping powers between 40-\mW{180} to shift the value of $f_p$. Any variation between the spectra beyond \kHz{1} is obscured by the noise demonstrating that there is a negligible influence from changes in the angular trap stiffness. It is important to note that the increased trapping does not cause any observable heating as evidenced by the near-identical spectra. An increase in temperature would reduce the viscosity and shift the velocity thermalisation corner at $f_i$ towards lower frequencies. 

This result can be exploited to allow for calibration-free measurements where the trap stiffness does not need to be known. This was verified in Fig. \ref{fig:CalibrationFree}B which shows the estimated values of $f_i$ determined from fitting with a calibrated trap (where $f_p$ is known) is equivalent to fitting with an extreme `uncalibrated' case assuming $f_p=0$. The calibration-free fits remain accurate provided that the sample time is shorter than $\tau_p$. The SNR is not affected by changes in the trap power because the angular position sensing comes from the independent beam beam allowing trapping with very low powers without loss of sensitivity. The optical potential is only required to confine the probe and control the equilibrium orientation. 

Ultrafast viscometry is an ideal application of ballistic tweezers to take advantage of short measurement times and calibration-free analysis. Viscosity, $\eta$, is usually defined in the long-time limit as a measure of a fluid's resistance to flow. At low Reynold's numbers, this is often linked to the diffusion rate or Stoke's drag of a particle in the fluid. At short times, the hydrodynamic interactions modify the drag on the probe such that it still depends on viscosity but is no longer linearly proportional to velocity. However, it can be determined from the fast thermalisation of angular velocity with $ \eta =2 \pi a^2 \rho f_i / 15$. It is important to reiterate that this result gives the same viscosity as defined in the long-time limit. As ballistic measurements have negligible influence from the optical trap, we have developed a rotational microviscometry technique decoupled from the optical potential and only requiring short measurement times to resolve. 

\begin{figure}[ht]
    \centering
    \includegraphics[width=0.9\linewidth]{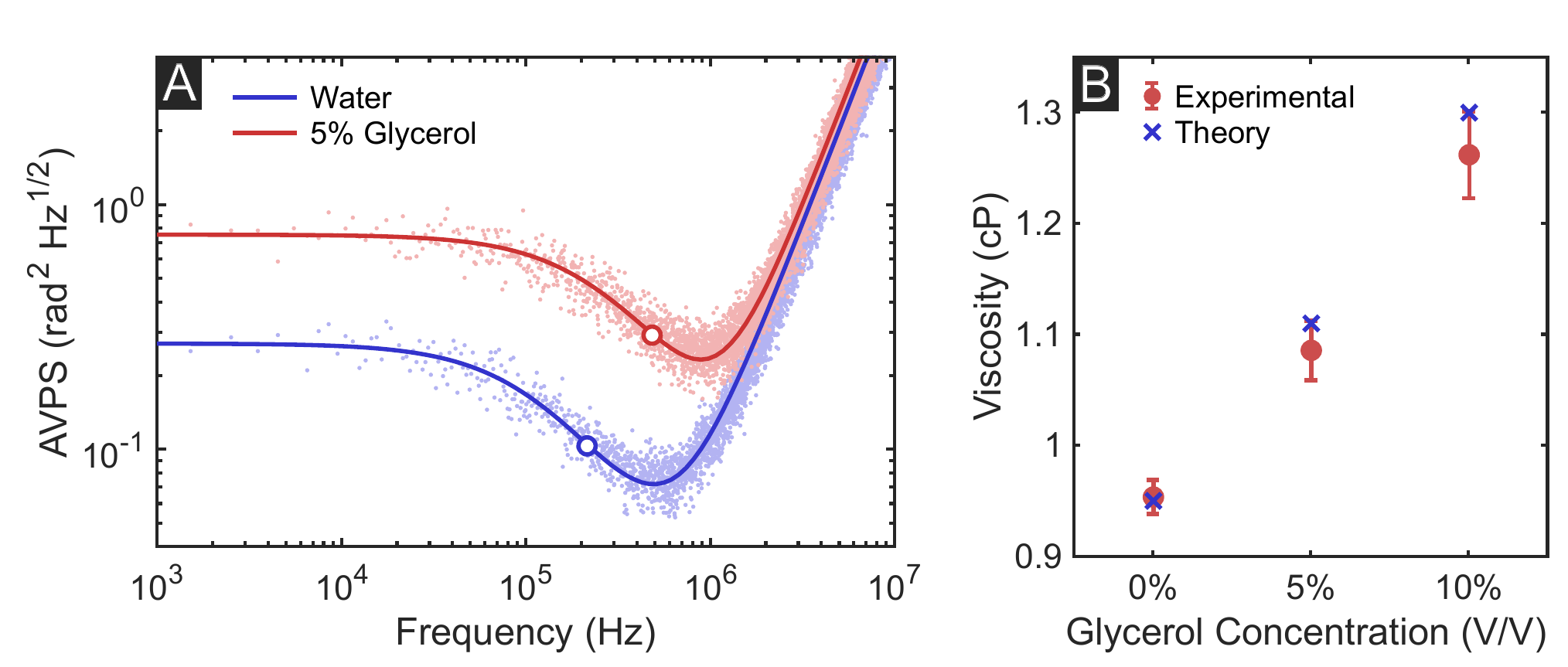}
    \caption{(A) Experimental AVPS of 0\% (blue) and 5\% glycerol aqueous (red) solutions and the corresponding $f_i$ values determined from fitting (circles). (B) Ballistic viscometry measurements of glycerol aqueous solutions (red) compared to literature values (blue crosses) \cite{Volk2018}. The error bars are propagated from the parameter estimation and the uncertainty in radius. }
    \label{fig:viscometry}
\end{figure}

We performed viscometry of water and aqueous glycerol solutions as proof-of-principle tests for rotational ballistic viscometry. Fig. \ref{fig:viscometry} shows two sample curves and compares the measured viscosity values to literature values \cite{Volk2018}. The AVPS curves shown in Fig. \ref{fig:viscometry}A agree with the expected model. The variations in $D$ and $f_i$ result from the different viscosities of the two samples and the probe sizes. The slight shift in the magnitude of the noise floor is due to differences in the optical power on the detector. There is excellent agreement for the estimated water viscosity and good agreement for the glycerol solutions as shown in Fig. \ref{fig:viscometry}B. The increased uncertainty was partially caused by the higher $f_i$ values that have lower SNR and the smaller radius of the ballistic probes. The systematic underestimation is likely a result of either a \degC{1} discrepancy or the hygroscopic behaviour of glycerol, which reduces its viscosity. The uncertainty in the radius dominates the measurement uncertainty of $\eta$ as $\Delta\eta\propto 2\Delta a$ and $a$ is seldom determined with better precision than $5\%$ in vaterite trapping experiments \cite{Watson2022}. However, this is still an improvement from previous ROTs experiments because those were performed in the diffusive regime where $\Delta \eta \propto 3\Delta a$. The SNR at $f_i$ determines the ease with which the velocity dissipation rate is resolved and thus also contributes to $\Delta \eta$. 

The technique can be extended to viscoelastic fluids, however, it is unlikely to capture the elastic component because that normally becomes significant on longer timescales. Instead, the improved sensitivity will aid in the accuracy of those measurements and can still provide information about the short-time particle--liquid interactions. Coupling the ballistic measurements with existing active microrheometry would provide an unprecedented measurement bandwidth and sensitivity to perform microrheometry.

\section{Conclusion}

Access to rotational ballistic tweezers improves the speed at which fluid properties are determined, allowing the observation of the transition from diffusive to ballistic rotational motion. We access this regime through improvements in our experimental apparatus that combine measurement basis optimization with reduced detector noise, significantly improving the SNR from previous experiments. In addition to the trapping beam, we incorporate a stable measurement beam which allows for large variations in the trapping power without reducing the SNR. The polarisation-sensitive technique is the optimal measurement basis for performing rotational ballistic measurements approaching the standard quantum limit, decoupled from the optical potential when performing ultrafast measurements. 

We demonstrated that our methodology is capable of resolving angular velocity thermalisation in liquids to access the ballistic regime allowing validation of hydrodynamic models and close-to-real-time measurements of viscosity. Employing this methodology to investigate systems out-of-equilibrium offers an exciting avenue of research. Applications into complex systems including studies of non-Newtonian fluids, non-equilibrium Brownian dynamics, biological processes, and those favouring rotational geometry, will greatly benefit from the improved sensitivity and fast measurement times of rotational ballistic tweezers measurements and allow for comparison with numerical predictions \cite{Watson2022, Garduno2015, Li2023}. Combining rotational and translational ballistic techniques and extending the analysis to non-spherical particles would provide an opportunity to investigate force-torque coupling. An attractive example of such a measurement would be to investigate the translational dynamics and corresponding force on the surrounding fluid of a micromachine driven by an optical torque. This would provide insight into the dynamics of diverse optically driven nano- and micromachines, microswimmers and other hydrodynamic models \cite{Felderhof2007, Zhang2019, Fouxon2019, Andren2021, Candelier2023, Viot2024}. Our enhanced measurement sensitivity and reduced measurement time have applications to biophysical measurements of cell--fluid interactions and in dynamic cellular environments where traditional approaches may be too slow and lack the spatial resolution to accurately capture the dynamics.

\section{Acknowledgements}
We acknowledge the funding from the Australian Research Council (DP230100675, DP220103812, DE230100972), the ARC CoE in Quantum Biotechnology (CE230100021), the Australian National Health and Medical Research Council (Ideas Grant 2012140), NIH (RM2018001458), and the Australian Government Research Training Program Scholarship. We acknowledge the Traditional Owners and their custodianship of the lands on which UQ operates. 

We declare no conflicts of interest. Data is available upon request.


\end{document}